\documentclass[10pt,twocolumn,floatfix,prb,superscriptaddress,]{revtex4-2}
\usepackage{amsmath,amssymb,amsthm,mathrsfs,amsfonts,dsfont,amstext,bbold} 
\usepackage{textcomp,pbox}
\usepackage[export]{adjustbox}
\usepackage{bm}
\usepackage{dcolumn,booktabs,url}
\usepackage[scaled]{helvet}
\usepackage{sansmath,gensymb}
\usepackage{tikz,graphicx,transparent,color}
\usepackage{multirow}
\usepackage[separate-uncertainty = true]{siunitx}
\usepackage{comment}
\usepackage{physics}
\usepackage{pdfpages}
\usepackage{xcolor}
\usepackage[english]{babel}

\makeatletter
\AtBeginDocument{\let\LS@rot\@undefined}
\makeatother

\usepackage[colorlinks=true]{hyperref}
\usepackage{graphicx}

\hypersetup{
	colorlinks   = true,
	citecolor    = red,
	linkcolor    = blue,
	urlcolor     = red     
}

\graphicspath{{./figs/}}

\newcommand{\euiso}[0]{$^{151}$Eu$^{3+}$:Y$_2$SiO$_5$}

\begin{document}
	
	\newcommand{\TitleName}{A readout-integrated time-bin qutrit analyzer for echo-based quantum memories}
	\title{\TitleName}
	
	\newcommand{\AffGeneve}{Département de Physique Appliqu\'{e}e, Universit\'e de Gen\`{e}ve, CH-1211 Gen\`{e}ve, Switzerland}
	\newcommand{\AffNice}{Université c\^{o}te d’Azur, CNRS, Institut de Physique de Nice (INPHYNI), UMR 7010, Parc Valrose, Nice Cedex 2, France}

	\author{Adrian Holz\"{a}pfel} 
	\affiliation{\AffGeneve{}}
	\author{Antonio Ortu} 
	\affiliation{\AffGeneve{}}
	\author{Mikael Afzelius} 
	\affiliation{\AffGeneve{}}

	\date{\today}
	
	\begin{abstract}
		We present a method to project time-bin qutrits stored in an echo-based quantum memory using several successive partial readouts of the memory. We demonstrate how this scheme can be used to implement projections onto a full set of mutually unbiased bases and, therefore, enables the characterization of arbitrary quantum states. Further, we study the integration of this protocol for the case of atomic frequency comb spin-wave storage by simulating the full storage process and performing a storage experiment with bright time-bin pulses in \euiso{}. In this context, a compound pulse for implementing partial readouts in quick succession is introduced and characterized.
	\end{abstract}
	
	\maketitle 
	
	%%%%%%%%%%%%%%%%%%%%%%%%%%%INTRODUCTION%%%%%%%%%%%%%%%%%%%%%%%%%%%%%%%%%%%%%%%%%
	%%%%%%%%%%%%%%%%%%%%%%%%%%%%%%%%%%%%%%%%%%%%%%%%%%%%%%%%%%%%%%%%%%%%%%%%%%%%%%
	
	\section{Introduction}
	Long distance quantum networks would allow for a plethora of applications ranging from unconditionally secure communication to distributed quantum simulation and computation~\cite{Kimble2008}. The prerequisite for long-distance links within such a network is the efficient mapping of information stored in photonic degrees of freedom to and from quantum memories~\cite{Briegel1998,Duan2001,Sangouard2011}.
	A common proposal for such networks is to encode information in two dimensional degrees of freedom, such that each flying photon carries one quantum bit (qubit) of information. Encoding in higher dimensional spaces, however, has been shown to have several advantages over the encoding in qubits, showing a higher resilience to noise~\cite{PhysRevLett.85.4418,PhysRevLett.104.060401,PhysRevX.9.041042} as well as improved secret-key rates in quantum key distribution~\cite{PhysRevA.87.062322,PhysRevLett.98.230501,PhysRevA.82.030301}.
	
	A suitable candidate for implementing high dimensional photonic quantum states are time-bin qudits. For these, entangled states can be readily generated~\cite{deRiedmatten2002creating,PhysRevA.69.050304}, and temporally multimode quantum memories can be used for their storage~\cite{Gundogan2015,Tiranov2016b,Tiranov2016c,ortu2021storage}. For time-bin qubits it has been demonstrated that in certain types of echo-based optical memories, projective measurements can be conveniently integrated into the readout process~\cite{Timoney2013,PhysRevLett.98.113601,Gundogan2013}. Specifically, two successive partial readouts can be used to interfere neighboring time-bins and, as such, allow for projection onto superposition states. This approach is equivalent to the use of an unbalanced Mach-Zehnder-interferometer (uMZI) to perform time-bin qubit projections~\cite{Gundogan2013}. While such a Mach-Zehnder analyzer (MZA) can be readily generalized to higher dimensions by increasing the amount of spatial modes of the uMZI~\cite{thew2004bell}, it is not immediately obvious how to implement an equivalent action with partial readouts of a quantum memory. For a partial readout analyzer (PRA) the atomic transitions that play the role equivalent to spatial modes in the MZA is principally limited to two.
	
	In this article we propose a time-bin qutrit analyser based on a succession of amplitude and phase-optimized 2-mode-beam-splitter-like interactions. It is shown that such an analyser can exceed the conventional time-bin qutrit analyser based on 3-mode beam splitters in efficiency. The implementation of this protocol requires several efficient partial readouts of the memory in close succession. To this end we propose a novel composite adiabatic pulse that can implement this action within the constraints of limited read/write power. A full qutrit storage and analysis scenario using the composite adiabatic pulses is simulated with Maxwell-Bloch simulations, based on the atomic frequency comb (AFC) spin-wave memory protocol. We also present an experimental AFC spin-wave implementation of the qutrit storage and analysis in a \euiso{}  crystal, in the regime of bright time-bin pulses.

	\section{Theory}\label{SEC:theory}

	A time-bin qudit is formed by bringing a photon into a superposition of $d$ possible arrival times, that we will represent as
	\begin{equation}\label{EQU:timebinqudits}
		\ket{\psi}= \sum_{i=0}^{d-1} a_i\ket{i}_t
	\end{equation}
	where $\ket{i+1}_t$ denotes a temporally localized wave packet that is delayed by time-bin width $\tau$ from the otherwise identical wave packet $\ket{i}_t$. This time-bin qudit can be measured in the canonical basis $\{\ket{i}_t\}_{i=0}^{d-1}$ simply by performing time resolved single photon detection. But a full characterization of the quantum state requires measurements with a complete set of complementary observables. For a time-bin qubit ($d=2$), for example, simple time resolved detection corresponds to a measurement of Pauli operator $\sigma_z$. For a full characterization we would like to also implement measurements of the Pauli operators $\sigma_x$ and $\sigma_y$. 
	
\begin{figure}[ht!]
		\includegraphics[width=0.75\columnwidth]{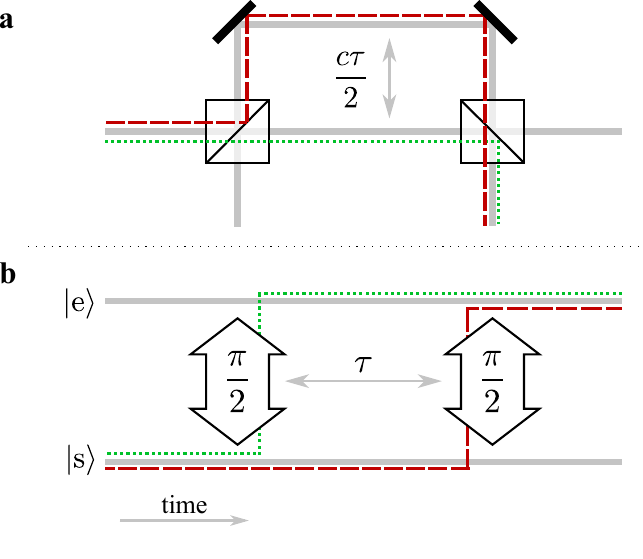}
		\caption{\label{FIG:twopulses}
		\textbf{Two equivalent methods for measuring superposition states of time-bin qubits.}
			(a) An uMZI with overall path difference $\Delta s=c\tau$ interferes adjacent time-bins with bin width $\tau$.
			(b) An equivalent projector can be implemented by means of two partial readouts of a quantum memory. As the time evolution in the storage state $\ket{s}$ can be considered frozen, coherence that is mapped to the excited state via the green, dotted path will rephase $\tau$ earlier than coherence following the red, dashed path. The equivalent paths in the uMZI are indicated in the same color scheme.
		}
	\end{figure}

	The MZA is a common method for implementing these measurements that relies solely on passive, linear optical components. It uses an uMZI to interfere neighboring time-bins, as illustrated in figure \ref{FIG:twopulses} (a). The uMZI consists of two arms of different length such that a photon taking the longer path through the device is delayed by $\tau$ relative to a photon taking the shorter path. Thus, the possibility of a photon arriving in the earlier time-bin but taking the long path through the device and a photon arriving late but taking the short path interfere at the output. By adjusting the phase between the two paths of the uMZI appropriately, we can realize projections onto states of the form $\frac{1}{\sqrt{2}}(\ket{0}_t+e^{\mathrm{i}\phi}\ket{1}_t)$ which in particular includes the eigenstates of $\sigma_x$ and $\sigma_y$. 
	
	We note that this MZA is only conclusive in 50$\%$ of the cases, as an early photon might as well take the short path, or a late one the long path, in which case no interference will take place. This issue can be avoided by replacing the first beam splitter with an active switch~\cite{PhysRevA.61.062308}. The losses of currently available switches, however, relativize the advantage of having no inconclusive results, such that passive devices are commonly used. In the following we will compare our method only to the conventional MZA with passive components.
	
\begin{figure}[ht!]
		\includegraphics[width=0.9\columnwidth]{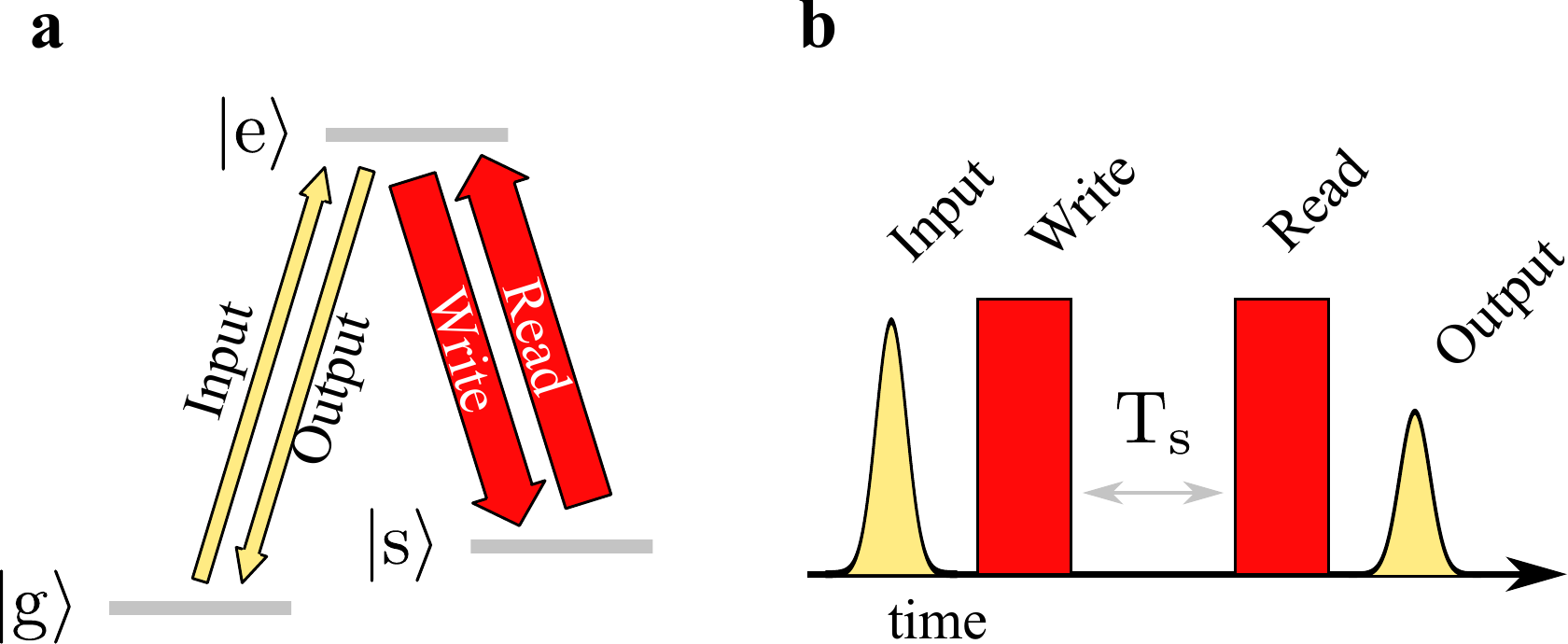}
		\caption{\label{FIG:genericmemory}
			\textbf{Energy structure (a) and temporal sequence (b) of quantum storage in a lambda system.}
			The incoming photon (input) is absorbed on transition $\ket{g}-\ket{e}$. An optical $\pi$-pulse (write) is mapping the resulting coherence onto the long-lived transition $\ket{g}-\ket{s}$. After a storage time of $\mathrm{T}_\mathrm{s}$ the coherence is mapped back onto $\ket{g}-\ket{e}$ by another optical $\pi$-pulse (read) and re-emitted as a flying photon (output). When considering the action of the read/write pulses, the memory can be treated as a two-level system with states $\ket{s}$ and $\ket{e}$, as $\ket{g}$ is unaffected by these pulses.
		}
	\end{figure}

	For on-demand and temporally multi-mode quantum memories like  CRIB/GEM~\cite{Alexander2007a,Hetet2008a,sabooni2020broadband}, ROSE~\cite{Damon2011,Bonarota2014} or AFC spin-wave storage~\cite{ortu2022multimode,Ortu2022b}, an equivalent action can be performed by means of two successive partial readouts of a stored time-bin qubit~\cite{Gundogan2013}. We consider a quantum memory consisting of a lambda system with a long-lived spin transition $\ket{g}-\ket{s}$ and two optical transition, $\ket{s}-\ket{e}$ and $\ket{g}-\ket{e}$. The latter interacts as interface that absorbs flying photon. The resulting coherence in the optical transition is then mapped onto the spin transition by means of an optical $\pi$-pulse on $\ket{s}-\ket{e}$ (write pulse). For retrieval, the coherence is mapped back onto $\ket{g}-\ket{e}$ by another $\pi$-pulse (read pulse) and, subsequently, re-emitted as a flying photon based on a process that is dependent on the particular protocol in use. This storage sequence is sketched out in figure~\ref{FIG:genericmemory}.
	
	For implementing a partial readout analyzer (PRA) the $\pi$-pulse is replaced by two $\pi/2$-pulses. If they are spaced apart by the width of one time-bin and the second partial readout occurs before the first emission of the memory, then their resulting action is completely equivalent to the MZA where $\ket{s}$ and $\ket{e}$ take on the role of the two spatial modes of the uMZI.  This is illustrated in figure \ref{FIG:twopulses} (b). The PRA is particularly useful for measuring time-bin qubits when the photons have a very long coherence time $>\SI{100}{\nano\second}$. The corresponding uMZI would require an optical path length difference of several meters or more between its two arms, such that stabilization becomes challenging. A quantum memory on the other hand is by its very nature capable of generating the required coherent delay.
	
	The MZA can be readily generalized to three dimensions. If each 2-mode beamsplitter is replaced by a 3-mode beamsplitter (tritter), any photon passing through the uMZI may take three different paths of different optical path length, such that all time-bins of a qutrit can interfere with each other~\cite{thew2004bell}. For the PRA, however, it is not immediately clear how to generalize the time-bin projection scheme to higher dimensions. As $\ket{s}$ and $\ket{g}$ are the equivalent to the spatial modes in the uMZI, the number of modes cannot be increased beyond two. We will demonstrate how despite this we can implement an efficient analyzer scheme. 
	
	\begin{figure}[ht]
		\includegraphics[width=0.75\columnwidth]{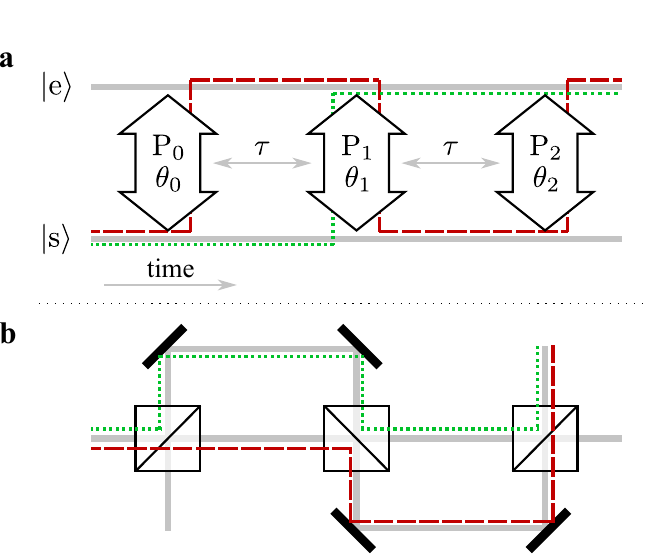}
		\caption{\label{FIG:threepulses}
			\textbf{Illustration of the readout-based qutrit analyzer.}
			(a) Each projection is implemented by applying three partial readouts with variable pulse area $P_i$ and phase $\theta_i$ with a relative delay of $\tau$ between each. There are two interfering quantum paths resulting in a delay of $\tau$ that are indicated by the dashed red and dotted green line respectively.
			(b) This linear optics circuit is equivalent to the partial-readout scheme above. Once more the two interfering quantum paths are indicated.
		}
	\end{figure} 
	
	Naively, one might expect that one can use three partial readouts of the memory to overlap three neighboring time-bins in complete analogy to the qubit PRA. However, while for the qubit case each generated delay inherits its phase from exactly one readout pulse, for the qutrit case two possible quantum paths through the interferometer lead to the same delay, as illustrated in figure \ref{FIG:threepulses}. One of these paths inherits the phases from all readout pulses. As consequence the phase and amplitude of this delay depends non-trivially on the phase and amplitude of all of the readout pulses.
	
	 We will use an effective model of the partial readouts to put this into quantitative terms. For this purpose we will describe the system as a tensor product $\mathbf{H_t}\otimes\mathbf{H_s}$, where $\mathbf{H_t}$ represents the time-bin space as introduced in equation \ref{EQU:timebinqudits} and  $\mathbf{H_s}$ represents the internal state of the memory, which can be either $\ket{s}$ or $\ket{e}$ and specifies whether a given coherence is transition $\ket{g}\leftrightarrow\ket{s}$ or $\ket{g}\leftrightarrow\ket{e}$. 
	 
	 We choose this notation to highlight the equivalence between the PRA and a 2-mode linear optical device, where $\ket{s}$ and $\ket{e}$ would correspond to two spatial modes. It should be noted, however, that for the PRA strictly speaking $\mathbf{H_t}$ describes coherences on either spin transition $\ket{g}\leftrightarrow\ket{s}$ or optical transition $\ket{g}\leftrightarrow\ket{e}$. The time-bin state that eventually will be emitted by the memory corresponds to whatever coherence has been mapped to $\ket{g}\leftrightarrow\ket{e}$ after all three partial readout pulses have been applied.
	
	The effect of an ideal readout pulse with an arbitrary phase and area on the $\ket{s}$-$\ket{e}$ transition can then be written as:
	\begin{equation}
		\mathbf{A}=\mathbf{Id}_t\otimes \begin{pmatrix}
			a & b\\
			-b^\star & a
		\end{pmatrix}_s
	\end{equation}
	with 
	
	$$a=\sqrt{1-P}$$
	$$b=e^{\mathrm{i}\theta}\sqrt{P}$$
	where $P$ is the transfer probability and $\theta$ the phase of the pulse. 
	
	Coherences on the spin transition can be considered frozen in their time evolution while coherences on the optical transition are experiencing rephasing. That means that the rephasing of any coherence stored in the spin transition is delayed by one time-bin if the system is left to freely evolve for a time $\tau$. We can model this evolution with the effective operator
	\begin{equation}
		\mathbf{B}=\mathbf{C}_t\otimes\dyad{s}{s}_s+\mathbf{Id}_t\otimes\dyad{e}{e}_s
	\end{equation}
	with \[\mathbf{C}_t= \sum_{i=0}^{\infty} \ket{i+1}\bra{i} \]
	
	Applying three readout pulses with free evolution of $\tau$ in-between them then performs the following unitary operation on the system.
	\begin{equation}
		\mathbf{U}=\mathbf{A_2BA_1BA_0}
	\end{equation}
	The element of this matrix that we are interested in is the one that maps from $\ket{g}\leftrightarrow\ket{e}$ to $\ket{g}\leftrightarrow\ket{e}$. We find the following expression
	\begin{equation}
		%a a2 c^2 Conjugate[b1] + a^2 c Conjugate[b2] + a a2 Conjugate[b3] + b2 c Conjugate[b1] Conjugate[b3]
		\begin{split}
		&\bra{e}\mathbf{U}\ket{s}=\\
		&-b_0^*a_1a_2\mathbf{Id}_t-a_0b_1^*a_3 \mathbf{C}_t+ b_0^* b_1 b_2^* \mathbf{C}_t - a_0a_1b_2^* \mathbf{C}_t^2
		\end{split}
	\end{equation}
	Ordering by experienced delay shows that a time-bin eigenstate entering the interferometer will exit as a superposition of three time-bins with complex amplitudes
	\begin{equation} 
		\begin{split}
			& \zeta_0=-a_0a_1b_2^* \\
			& \zeta_1=-a_0b_1^*a_2 +b_0^* b_1 b_2^* \\
			& \zeta_2=-b_0^*a_1a_2
		\end{split}
	\end{equation}
	
	\begin{table*}[ht!]
 		\centering
 		\begin{tabular}{|c|c|c|c||c|c|c|c|c|c|}
 			\hline
 			\textbf{Basis}	&  $\phi_0$		&  $\phi_1$	& 	$\phi_2$  &	 $P_{0,2}$	& $P_{1}$	& ${\theta_0}$ & ${\theta_1}$ & ${\theta_2}$ & $\eta$ \\
 			\hline
 			\hline
 			MUB 1	&  0 &  0	& 0 & $(3-\sqrt{3})/6$ & $1/3$ & 0	&  0	& 	0  & $1/3$ \\
 				&  0 &  $-2/3\pi$	& $2/3\pi$ & $(3-\sqrt{3})/6$ & $1/3$ & $2/3\pi$	& $-2/3\pi$ &  0  & $1/3$ \\
 				&  0 &  $2/3\pi$	& $-2/3\pi$& $(3-\sqrt{3})/6$ & $1/3$ & $-2/3\pi$	& $2/3\pi$ &  0   & $1/3$ \\
 				\hline
 			MUB 2	&  0 &  0	& $-2/3\pi$& 0.276 & 0.286 & $-2/3\pi$	&  -0.388	& 	0   & 0.429 \\
 			&  0 &  $-2/3\pi$	& 0& 0.276 & 0.286 & 0  & -2.482  &  0  & 0.429 \\
 			&   $-2/3\pi$ &  0	& 0 & 0.276 & 0.286 & 0	&  -0.388	& 	$-2/3\pi$  & 0.429 \\
 			\hline
 			MUB 3	&  0 &  0	& $2/3\pi$& 0.276 & 0.286 & $2/3\pi$	&  0.388	& 	0 & 0.429 \\
 			&  0 &  $2/3\pi$	& 0 & 0.276 & 0.286 & 0  & 2.482  &  0  & 0.429 \\
 			&   $2/3\pi$ &  0	& 0 & 0.276 & 0.286 & 0	&  0.388	& 	$2/3\pi$  & 0.429 \\
 			\hline
 			\hline
 			Optimal 	&  0 &   $\pi/2$	& 0 & 0.5 & 0.2 &  0 &   $\pi/2$	& 0 & $3/5$ \\
 			basis &  $-2/3\pi$ &   $\pi/2$	& $2/3\pi$& 0.5 & 0.2 &  $2/3\pi$ &   $\pi/2$	& $-2/3\pi$  & 3/5 \\
 			&  $2/3\pi$ &   $\pi/2$	& $-2/3\pi$ & 0.5 & 0.2&  $-2/3\pi$ &   $\pi/2$	& $2/3\pi$  & 3/5 \\
 			\hline
 			
 		\end{tabular}
 		\caption{\textbf{Pulse parameters and theoretically predicted efficiency for three MUBs and the optimal basis.}
 			Each basis consists of three projectors of the form $(e^{\mathrm{i}\phi_0}\ket{0}+e^{\mathrm{i}\phi_1}\ket{1}+e^{\mathrm{i}\phi_2}\ket{2})/\sqrt{3}$. The three readout pulses to implement a given projector have the transfer probabilities $P_i$ and the phases $\theta_i$, the resulting projection an efficiency of $\eta$.}
 		\label{TAB:MUBparameters}
 		\end{table*}
	
	To implement a given projection, it needs to be ensured that these three amplitudes have the appropriate magnitude and phase. Ideally, we would like to construct projections onto a complete set of mutually unbiased bases (MUBs)~\cite{PhysRevLett.85.3313}. For a $d$-dimensional Hilbert space this is a set of $d+1$ orthonormal bases
	\begin{equation}
	    \mathcal{S}_\mathrm{MUB}=\{\{\ket{\psi_i^l}\}_{i=0}^{d-1}\}_{m=0}^{d}
	\end{equation}
	 where the $i$ indexes the states within a basis and $m$ the different bases, such that the inner product of any two vectors from two different bases has the same magnitude
	\begin{equation}
	    m\neq n:~|\bra{\psi_i^m}\ket{\psi_j^n}|^2=\frac{1}{d}
	\end{equation}
	When such a set exists, it allows to fully characterize arbitrary states with the least amount of redundancy in-between measurements and can be seen as the higher dimensional generalization of characterizing a qubit by measuring with all three Pauli operators~\cite{Filippov_2011}. In three dimensions any vector of a basis that is mutually unbiased to the canonical one can be written in the form
	\begin{equation}
		\ket{\psi}=\frac{1}{\sqrt{3}}(e^{\mathrm{i}\phi_0}\ket{0}_t +e^{\mathrm{i}\phi_1}\ket{1}_t+e^{\mathrm{i}\phi_2}\ket{2}_t)
	\end{equation}
	
	As it will be shown, projection onto this state with a PRA scheme can be achieved by implementing the following mapping
	\begin{equation}\label{EQU:mapping}
		\begin{split}
			&\mathbf{U}_\psi\ket{n}_t\ket{s}=\\ &\sqrt{\frac{\eta}{3}}(e^{-\mathrm{i}\phi_2}\ket{n}_t +e^{-\mathrm{i}\phi_1}\ket{n+1}_t+e^{-\mathrm{i}\phi_0}\ket{n+2}_t)\ket{e}\\
			+&\sqrt{1-\eta}\ket{\psi}_t\ket{s}
		\end{split}
	\end{equation}
	where $\eta\in[0,1]$ and $\ket{\psi}_t$ symbolizes all coherence that has not been mapped onto $\ket{g}\leftrightarrow\ket{e}$ after all three pulses have been applied.
	
	Once we have implemented this mapping, we can perform the desired projection by time resolved detection after the memory.
	\begin{equation}
		\begin{split}
			& \ket{\phi}=\alpha\ket{0}_t+\beta\ket{1}_t+\gamma\ket{2}_t;\\ 
			&|\bra{e}_s\bra{2}_t\mathbf{U}_\psi\ket{\phi}\ket{s}|^2=\eta|\bra{\psi}\ket{\phi}|^2
		\end{split}
	\end{equation}
	In other words, after applying $\mathbf{U}_\psi$, the detection probability of a photon in time-bin $\ket{2}_t$ is proportional to the overlap of the stored state $\ket{\phi}$ and $\ket{\psi}$, the MUB state onto which we want to project the stored state. It reaches a maximum of $\eta$ when the two states are identical. We will refer to $\eta$ as the efficiency of the analyzer.
 	
 	A step-by-step procedure on how to find the optimal choice for all parameters for a given projector is presented in detail in the appendix \ref{SEC:pra_parameters}. A summary of optimal parameters for a complete set of bases that are mutually unbiased to each other and the canonical basis (MUBs) is found in table \ref{TAB:MUBparameters} (upper half). We note that the efficiency of the projections is dependent on the exact choice of the phases. This is a consequence of the aforementioned interference of two quantum paths contributing to the same delay. In its presence the magnitude of this delay and its relative phase with regard to the other two delays are no longer independent.

 	Besides this set of observables, we have investigated one more measurement basis. Instead of deciding for a specific projection and then finding the read pulse phases that implement it, we choose the phases such that they allow for fully constructive interference of the two quantum paths that result in the same delay. Then, the efficiency $\eta$ of the resulting analyzer is maximal. Constructive interference will occur whenever $\phi_0=-\phi_2$ and $\phi_1=\pm\pi/2$. An example for a measurement basis that can be constructed  from vectors of this form is appended to table \ref{TAB:MUBparameters}. The efficiency of this configuration is not only the optimal one for the proposed three-partial-readouts scenario, but indeed it can be shown that there can be no unitary black box that implements this action in a more efficient manner, as demonstrated in the appendix (\ref{SEC:blackbox}).  We further note that its efficiency is considerably higher than the value of $1/3$ that is reached for projections in the usual 3-mode analyzer scheme.

	\section{Implementation in an AFC spin-wave memory}
	To confirm the experimental feasibility of our theoretical considerations, we have applied them to the case of atomic frequency comb (AFC) spin-wave storage. We have studied its performance both within the scope of a Maxwell-Bloch simulation of the storage and readout sequence, as well as in a storage experiment with bright time-bin pulses.
	
	\begin{figure*}[ht!]
		\includegraphics[width=\textwidth]{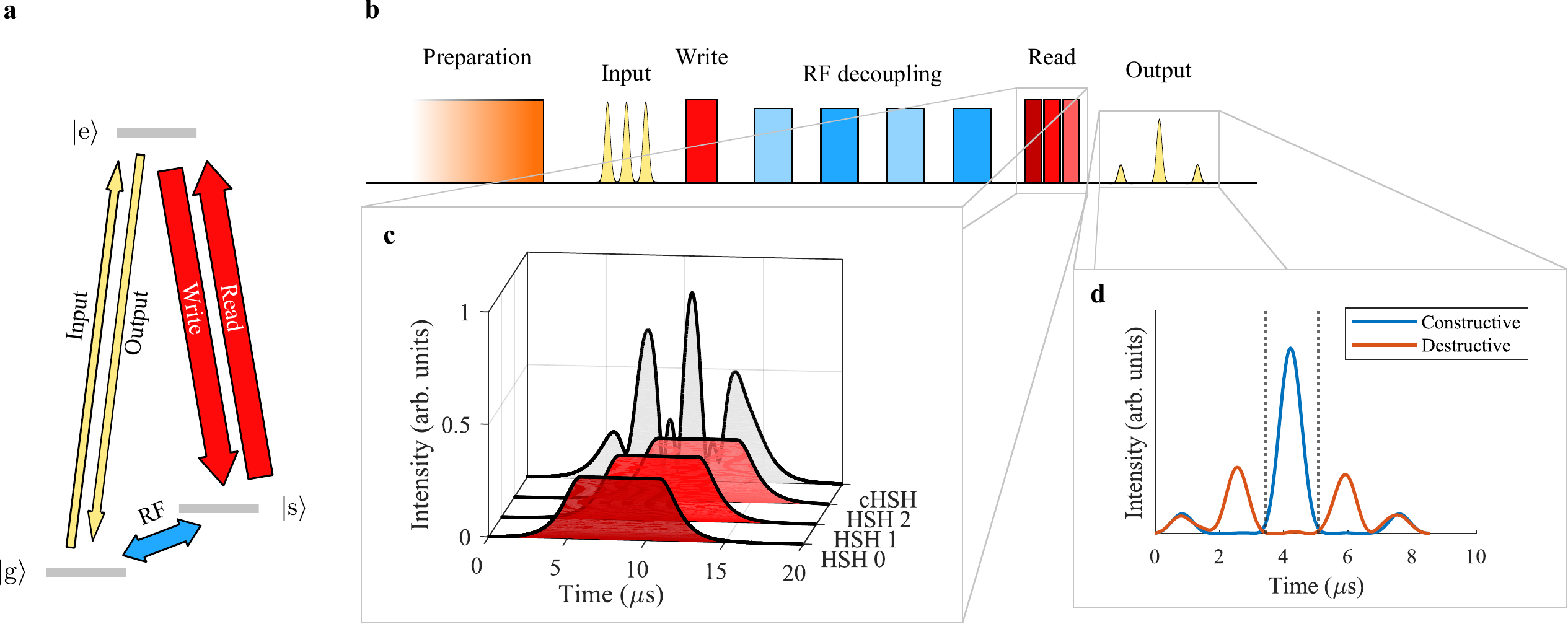}
		\caption{\label{FIG:sequence} 
			\textbf{AFC spin-wave storage with integrated PRA scheme.} (a) Level structure of the memory. (b) Temporal experimental sequence. (c) For the memory readout, we send a composite pulse that consists out of three overlapping HSH pulses (envelopes shown in red). An example for the envelope of the resulting composite HSH (cHSH) pulse is shown in gray. (d) Examples of simulated qutrit interference traces for constructive (blue) and destructive (red) interference. The interference bin is marked with dotted vertical lines.}
	\end{figure*} 
	
	The implementation of an AFC spin-wave memory requires a lambda-system with a long-lived spin transition $\ket{g}-\ket{s}$ and an optical transition $\ket{g}-\ket{e}$. The optical transition has to be inhomogeneously broadened, with the inhomogeneous linewidth exceeding the homogeneous one by far. Then, an AFC can be prepared by frequency-selective optical pumping ions from $\ket{g}$ into an auxiliary state~\cite{Afzelius2009a}. If a comb of periodicity $\Delta$ absorbs a train of pulses, the resulting coherence will rephase after a storage time of $T_\mathrm{AFC}=1/\Delta$ resulting in coherent re-emission. In order to allow for on-demand, long-duration storage, a write pulse on transition $\ket{e}-\ket{s}$ is applied before the rephasal has occurred. This pulse converts the optical coherence into a spin coherence on transition $\ket{g}-\ket{s}$. To preserve this coherence, several radio-frequency (RF) pulses may be applied to compensate for dephasing stemming from an inhomogeneous broadening of the spin transition and to dynamically decouple from the environment~\cite{Holzapfel2020}. The stored light is retrieved by applying a readout pulse on $\ket{e}-\ket{s}$ to map the coherence back onto $\ket{g}-\ket{e}$. There, it will rephase due to the AFC and, finally, be re-emitted. This experimental sequence is illustrated in figure~\ref{FIG:sequence} (a) and (b).
	
	For implementing the PRA scheme, the readout pulse is replaced with three partial readouts as described in section \ref{SEC:theory}. Here, a difficulty arises from experimental constraints. Efficient multimode AFC storage requires long optical lifetime and, therefore, typically is implemented in systems with low optical oscillator strength. This means that the optical Rabi frequency is often small in comparison to the bandwidth of the memory. Accordingly, efficient write/read pulses can only be implemented with an adiabatic profile. If the length of the resulting read pulse exceeds the width of the stored time-bins, then the partial readouts cannot be performed as a sequence of such adiabatic pulses. Deformation-less retrieval of the stored pulses, however, can only be ensured if rate and range of the frequency chirp of each read pulse are matched to those of the write pulse~\cite{Minar2010}.
	
	To perform two successive partial readouts despite these constraints we previously have proposed and implemented a composite pulse for analyzing time-bin qubits~\cite{ortu2021storage}. We extend this approach to the three partial readouts that are necessary to analyze qutrits and take the opportunity to have a closer look at its performance in a Maxwell-Bloch simulation. The composite pulse is formed by overlapping several adiabatic pulses with a time delay of $\tau$ with respect to each other, such that the resulting electric field is given by the linear combination of the individual adiabatic profiles $f_\mathrm{adiab}(t)$
	\[E_\mathrm{comp}(t)= \sum_{n=0}^2 E_ie^{\mathrm{i}\theta_i}f_\mathrm{adiab}(t+n\tau) \]
	where $\theta_n$ can be set directly according to table \ref{TAB:MUBparameters}, while the field magnitude $E_n$ has to be chosen such that the $n$-th partial readout has a total transfer probability of $P_n$ as specified by the table. In this work we have combined three hyperbolic-square-hyperbolic (HSH) pulses~\cite{Tian2011a} to form a composite HSH (cHSH) pulse, as illustrated in figure~\ref{FIG:sequence} (c). The HSH profile allows for a fast and uniform adiabatic transfer over a large bandwidth. It consists of a central region with constant Rabi frequency and chirp rate that is flanked by a smooth ramp up and ramp down of the field with a hyperbolic secant profile~\cite{Seze2005}. The pulse shape is fully characterized by its total chirp range $\Gamma$, duration of ramp up/ramp down $T_H$, duration of region with constant Rabi frequency $T_S$ and the duration of the truncation window $T_C$ to which the pulse is confined.
	
	\subsection{Maxwell-Bloch simulation}
	The storage process is simulated by numerically solving the semi-classical Maxwell-Bloch (MB) equations for two classical fields, the input/output mode and the write/read mode interacting with the lambda system of the memory. These fields are assumed to be spatially one-dimensional, forward propagating, with a slowly varying envelope. The specific implementation of this 3-level MB simulator is described in Refs. \cite{Minar2010b,Afzelius2010}. Our main interest lies in observing whether the predicted efficiencies for PRA analyzers can be reached when using cHSH pulses with realistic pulse parameters. To this end we made several simplifications, namely we neglected inhomogeneous broadening on the spin transition and we did not include any population decay or dephasing terms in the Bloch equations. After the application of the write pulse, any remaining optical coherences were set to zero to take into account that in actual experiments the spin storage time is typically much longer than the coherence time of the optical transitions. This also avoids interferences due to remaining optical coherences during the memory readout phase.
	
	Before implementing the PRA projectors, we have simulated a regular storage experiment as a reference for efficiency. The memory input is a train of three Gaussian pulses that are spaced apart by $\tau=\SI{1.67}{\micro\second}$ and have a full width at half maximum (FWHM) in intensity of $\tau_\mathrm{in}=\tau/2.38$ with a hard cutoff duration of $\tau$~\cite{Ortu2022b}. The optical transition is shaped to an AFC with optical depth $d=4$, bandwidth $\mathrm{BW}=\SI{4}{\mega\hertz}$ and finesse $F=\pi/\left[\arctan(2\pi/d)\right]$~\cite{Bonarota2010a}.
	Finally, write and read operations are performed by applying an HSH pulse with $\Gamma/(2\pi)=\SI{1.5}{\mega\hertz}$, $T_S=\SI{6}{\micro\second}$, $T_H=T_S/2$ and truncation window $T_C=\SI{12}{\micro\second}$. The peak Rabi frequency was set to $\Omega/(2\pi)=\SI{350}{\kilo\hertz}$. With these parameters the simulation yields a total AFC spin-wave efficiency of $\eta_0=30.3\%$. This efficiency is almost exclusively limited by the theoretical efficiency of the AFC for the given finite optical depth, which is $32.1\%$~\cite{Afzelius2009a}.

	The PRA method requires precise setting  of the phase and transfer probability of each component of the cHSH pulse. Therefore, we have simulated the storage efficiency with a readout of varying amplitude and directly recorded the relationship between amplitude and transfer probability. By inverting and interpolating the recorded relationship we can ensure that each partial readout has precisely the intended area given in table \ref{TAB:MUBparameters}. An example trace of the simulated memory output of the PRA can be seen in figure \ref{FIG:sequence} (d). 
	
	To characterize measurements in the four bases from table \ref{TAB:MUBparameters} we take the following approach. For each basis $\{\ket{\psi_i},i\in\{0,1,2\}\}$ we simulate storage and readout for all possible combinations of input $\ket{\psi_i}_\mathrm{inp}$ and PRA projection $\ket{\psi_j}_\mathrm{ana}$.
	Then we can determine the average fidelity of measurements in this basis, so the fidelity between the target state and the state that is actually projected onto
	\begin{equation}
	    F=\frac{\sum_i|_\mathrm{ana}\bra{\psi_i}\ket{\psi_i}_\mathrm{inp}|^2}{\sum_{ij}|_\mathrm{ana}\bra{\psi_j}\ket{\psi_i}_\mathrm{inp}|^2}
	\end{equation}
	We also record the average efficiency $\eta$ of each basis as a percentage of $\eta_0$. The results are listed in table \ref{TAB:MUBsimulation}.
	
% 	Instead of relying on the analytical expression for the approximate pulse area of an HSH pulse~\cite{Businger2019}, we have decided to calibrate the pulse area by recording the efficiency of the storage process as a function of the amplitude of the readout pulse. By inverting and interpolating this relationship we can ensure that every partial readout has precisely the intended area. Then, we may construct the cHSH pulses with phases and areas as suggested by the theory. An example trace of the simulated memory output can be seen in figure \ref{FIG:sequence} For each of the four introduced bases we record the overlap of every projector with every basis state. From this we can estimate a fidelity for each observable. Further we report the average efficiency $\eta$ of the projections of each basis as a percentage of $\eta_0$. The results are listed in table \ref{TAB:MUBsimulation}.
	
	\begin{table}
 		\centering
 		\begin{tabular}{|c||c|c|c|}
 			%{$T_s \text{ (ms)}$}	& {$\mu_\text{in}$}		& {$\eta \,(\%)$}	& {SNR} 		&{$\mu_1$}\\
 			\hline
 			\textbf{Basis}	&  $\eta$ (predicted) &  $\eta$ (simulation)	& 	$F$ \\
 			\hline
 			\hline
 			MUB 1	&  1/3 &  33.7$\%$	& 99.1$\%$ \\
 			\hline
 			MUB 2	&  42.9$\%$ &  42.5$\%$	& 98.5$\%$ \\
 			\hline
 			MUB 3	&  42.9$\%$ &  42.9$\%$	& 98.0$\%$ \\
 			\hline
 			Opt. basis &  3/5&  61.9$\%$	& 97.7$\%$ \\
 			\hline
 		\end{tabular}
 		\caption{\textbf{Efficiency and fidelity of PRA qutrit projections as observed in the Maxwell-Bloch simulation.}
 		For each basis we report the average fidelity of projections in this basis and with the efficiency of the simulated analyzer with the one calculated in section \ref{SEC:theory}.}
 		\label{TAB:MUBsimulation}
 		\end{table}
	
    As seen there, the simulation shows fidelities that are close to unity and an excellent agreement between theoretically predicted and simulated projection efficiency. That the theoretical upper bounds for the analyzer efficiency can be fully saturated also confirms that partial readouts grouped together in a composite pulse can retain their original pulse area, without cross-talk affecting their performance.

	\subsection{Experimental implementation}
	
\begin{figure}[ht!]
		\includegraphics[width=0.95\columnwidth]{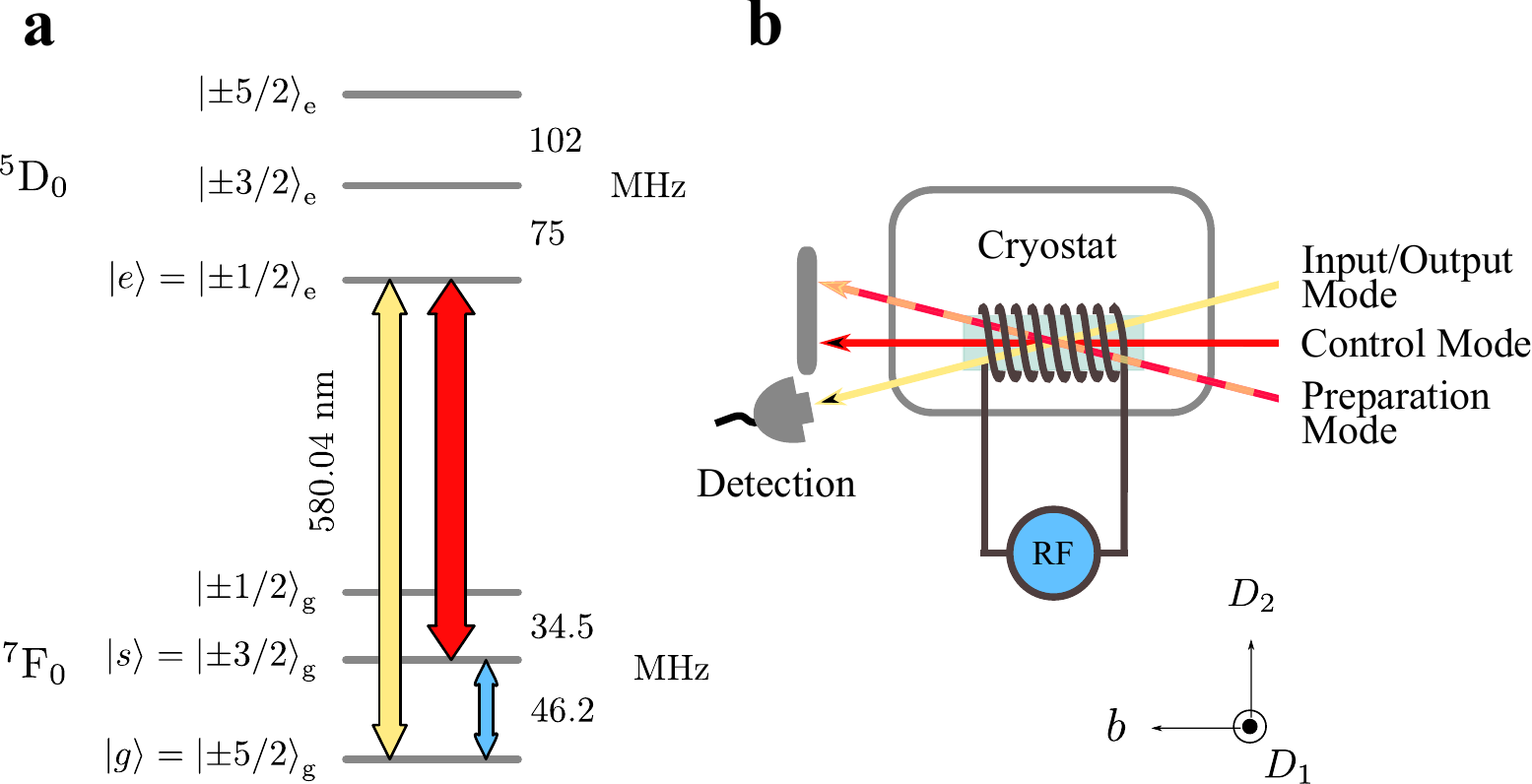}
		\caption{\label{FIG:setup} 
		\textbf{Simplified schematics of the AFC spin-wave experiment in \euiso{}.}
			(a) Relevant atomic energy structure and transitions used in the memory protocol. (b) Simplified sketch of experimental setup}
	\end{figure}

	Finally, we have implemented our method in an AFC spin-wave experiment where we have stored qutrits encoded onto bright time-bin pulses. As platform we have chosen a \euiso{} crystal with a doping concentration of 1000 ppm. The Y$_2$SiO$_5$ forms a biaxial crystal with polarization eigenaxes D$_1$, D$_2$, b \cite{Li1992}. The crystal is cut to a cuboid along these axes, with all laser beams traveling along the b-axis with their polarization oriented in D$_1$-direction such that absorption is maximized. Its temperature is kept at \SI{4}{\kelvin} using a closed-cycle helium cryostat. The relevant level structure is shown in figure \ref*{FIG:setup} (a). It consists of a ground and excited state that are connected by an optical transition of approximately \SI{580.04}{\nano\meter}.  Due to the Eu ion's nuclear spin of $5/2$, the quadrupolar interaction separates both the ground and the excited state into three doublets each, that are tens of MHz apart from each other. As the inhomogeneous linewidth of the optical transition is of the order of GHz, we need to apply a class-cleaning procedure\cite{Jobez2016,Lauritzen12} before every experimental sequence to ensure that we can optically address individual transitions. After this, we can define our lambda system as indicated in figure \ref*{FIG:setup} (a). 
	
	All optical fields are generated from a tunable external cavity diode laser at $\SI{1160}{\nano\metre}$ that is frequency locked onto a high-finesse cavity, amplified and finally frequency doubled~\cite{Jobez2014}. The resulting $\SI{580}{\nano\metre}$ beam is distributed onto three acousto-optic-modulators that supply three spatial modes for crystal preparation, write/read and input, respectively. Manipulation of the spin transition $\ket{g}-\ket{s}$ is performed with a coil connected to a resonance circuit~\cite{Ortu2022} that is driven by a $\SI{100}{\watt}$ RF-amplifier. This experimental setup is illustrated in figure \ref*{FIG:setup} (b). 
	
	For the storage experiment, we prepare an AFC with a bandwidth of \SI{3}{\mega\hertz} and an inverse periodicity of $1/\Delta=\SI{25}{\micro\second}$. Then, we send a train of three bright Gaussian pulses that form our qutrit with a time-bin width of $\tau=\SI{1.65}{\micro\second}$. Each pulse is truncated to the time-bin width and has a FWHM of $\tau_\mathrm{in}\tau/2.38$~\cite{Ortu2022b}. To map the coherence onto the spin transition, we use an HSH write pulse with $\Gamma/(2\pi)=\SI{1.5}{\mega\hertz}$, $T_H=\SI{1.65}{\micro\second}$  and a square duration of $T_H=\SI{7.5}{\micro\second}$. The pulse is truncated to a window of $T_C=\SI{15}{\micro\second}$. For its peak Rabi frequency, we measure approximately \SI{250}{\kilo\hertz}. After being mapped to the spin transition, the coherence is stored there for \SI{20}{\milli\second}. To preserve coherence, we perform a single XY4 decoupling sequence~\cite{GULLION1990479}. Additionally, during the whole experiment we apply a static magnetic field of $\SI{1.35}{\milli\tesla}$ along the $D_1$-axis of the crystal which has been shown to increase the coherence time of the spin transition significantly~\cite{Etesse2021,ortu2021storage}. A more detailed description of this setup and a characterization of its performance as a memory platform can be found in Ref.~\cite{ortu2021storage}.
	
	\begin{figure}[ht!]
		\includegraphics[width=\columnwidth]{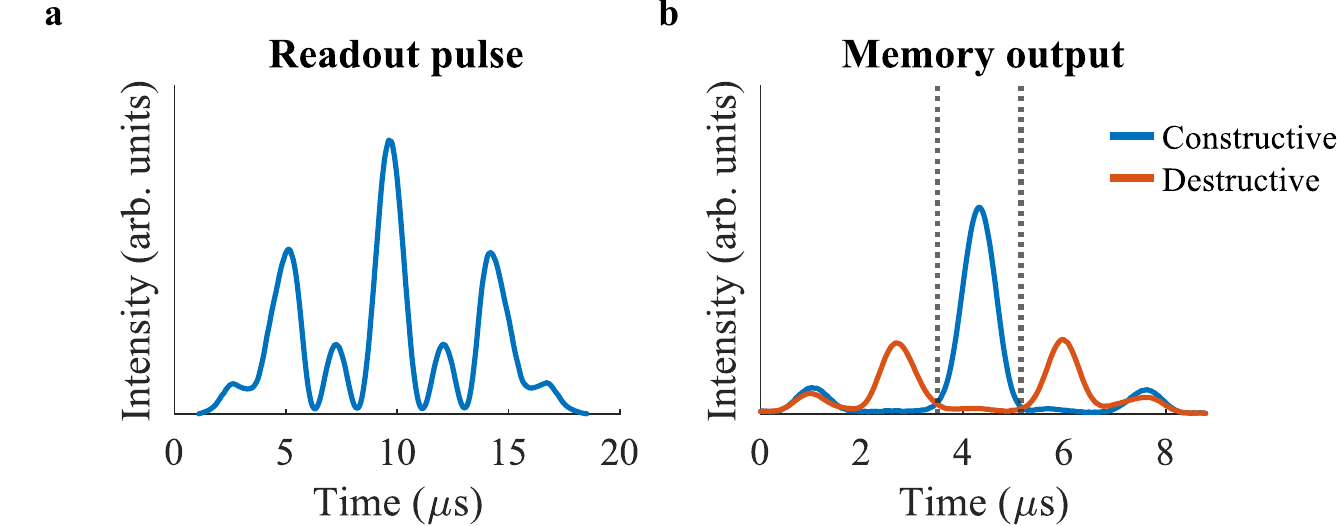}
		\caption{\label{FIG:examp_traces} 
			\textbf{Experimental traces} of readout (a) and memory output (b) as recorded by the photodetector.}
	\end{figure} 
	
	For the retrieval of the coherence we finally apply a cHSH pulse that implements the desired PRA. The three partial readouts are identical to the write pulse, except for their phase and amplitude. While the phases can be set directly according to our calculations, the optimal amplitudes are found by manual optimization. Examples of a cHSH pulse and the resulting qutrit interference pattern at the output are shown in figure~\ref{FIG:examp_traces}.
	
		\begin{figure}[ht!]
		\includegraphics[width=\columnwidth]{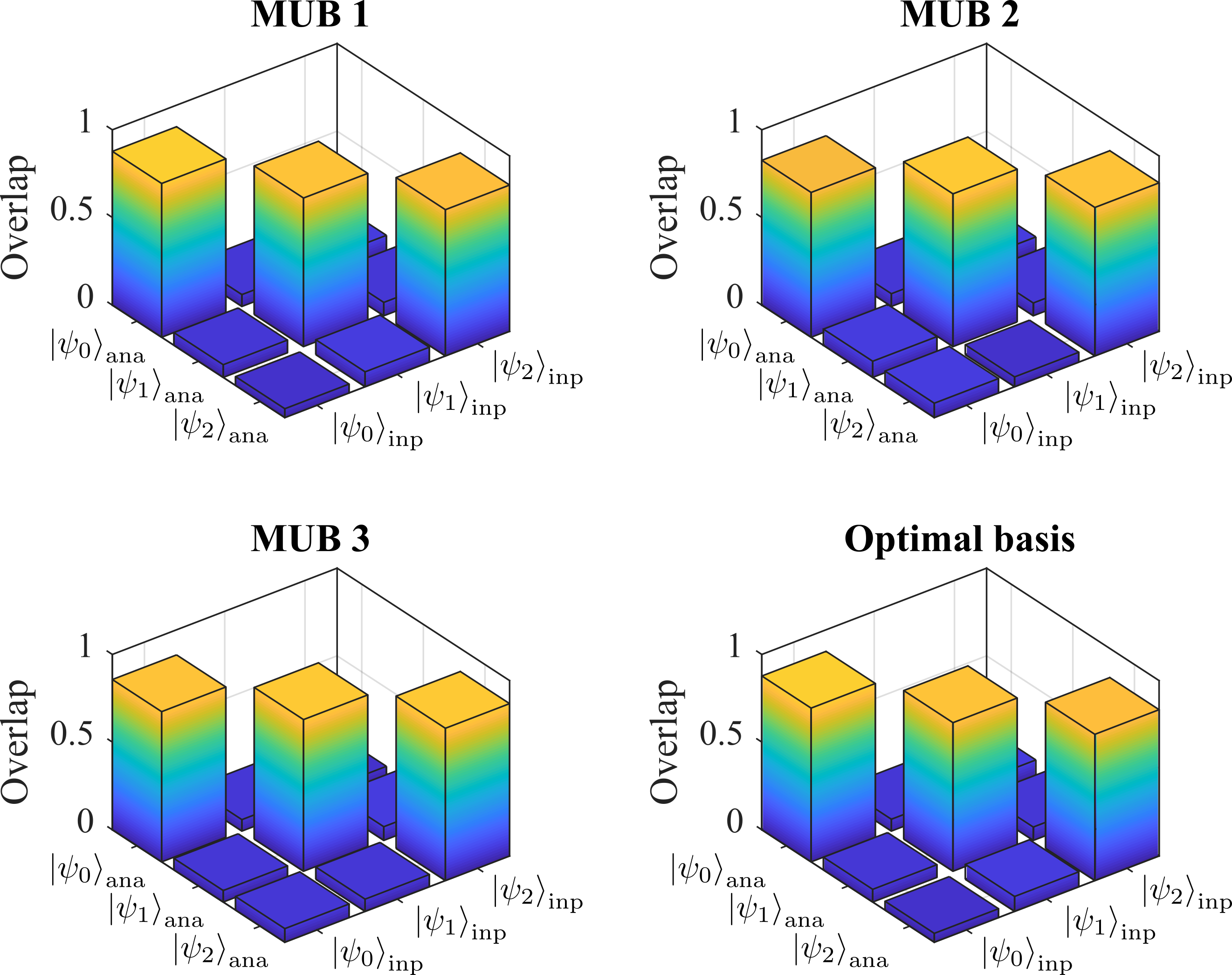}
		\caption{\label{FIG:fidelities} 
			\textbf{Overlap matrices of the four investigated bases as measured in storage with bright time-bin pulses.}
			For each basis PRA projections of each combination of input and analyzer setting are performed and their overlap recorded. The diagonal elements correspond to the fidelity of the respective projection.}
	\end{figure} 
	
\begin{figure}[ht!]
	\includegraphics[width=\columnwidth]{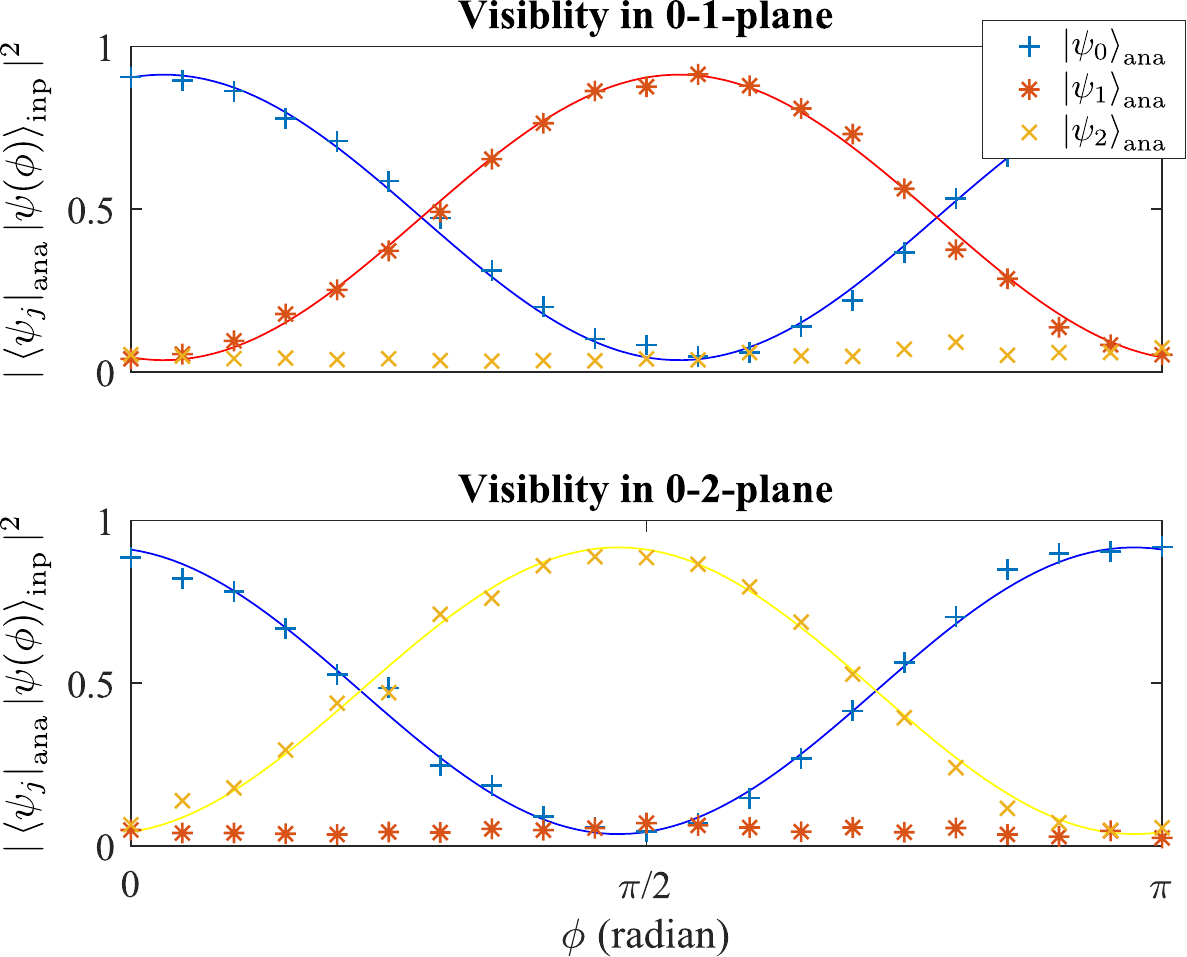}
	\caption{\label{FIG:visibilities} 
		\textbf{Visiblity measurement for the optimal basis in two different subspaces.}
		 The overlap with with three basis states $\ket{\psi}_\mathrm{ana}$ is recorded when the input is rotated in the 0-1-plane of the basis where\\
		 $\ket{\psi(\phi)}_\mathrm{inp}=\cos{\phi}\ket{\psi_0}_\mathrm{inp}+\sin{\phi}\ket{\psi_1}_\mathrm{inp}$
		 and in the 0-2-plane where\\
		 $\ket{\psi(\phi)}_\mathrm{inp}=\cos{\phi}\ket{\psi_0}_\mathrm{inp}+\sin{\phi}\ket{\psi_2}_\mathrm{inp}$.\\ The expectation value of the two projectors within the respective subspace is fitted with a sinusoidal model.
		 }
\end{figure}

	As before in the simulation, we have characterized the fidelity of the projectors by applying each of them to every vector of their respective basis. The resulting overlap matrices are shown in figure \ref{FIG:fidelities}. The observed average fidelity of the projectors is $F=(85\pm2)\%$. For sufficiently intense input pulses the spinwave storage is expected to essentially not affect the fidelity of the stored state because the noise generated by the storage process is negligible compared to the intensity of the retrieved light. Consequently, we can attribute any decrease in fidelity to imperfections of the qutrit measurement process. In order to determine whether the fidelity is limited by a unitary rotation of the analyzer from its nominal axis of projection, we have recorded visibility curves for one of the bases. For this purpose, we have measured the overlap for all three projectors while rotating the input state in two different sub-spaces. The resulting curves are shown in figure \ref{FIG:visibilities}. For both sub-spaces we observe that the projector is rotated by approximately $10^{\circ}$ from its nominal orientation. We estimate that the fidelity reported in the previous measurement could be increased by $3\%$ if this unitary error was compensated.

	\section{Conclusion}
	%shown how qutrits projection with two mode device
	%
	
	We have demonstrated how several partial readouts of a temporally multi-mode quantum memory can be used to project stored time-bin qutrits onto superposition states. The explicitly constructed projections onto a complete set of MUBs allow for the efficient characterization of arbitrary time-bin qutrit states. The Maxwell-Bloch simulation and storage experiment with bright pulses demonstrate that AFC spin-wave storage in conjunction with a compound pulse for the memory readout implement this measurement scheme in a way that can saturate the theoretically derived bound for projection efficiency in an ideal scenario and enables precise state characterization even in presence of experimental imperfections. \\
	The unitary operation that is implemented by the PRA scheme can also be realized as a linear optical device using only 2-mode beam splitters. This could be of interest since this device outperforms the 3-mode-splitter-based approach to qutrit measurement in efficiency, though at the price of implementing just a single projection at a time instead of up to three simultaneously. This trade-off might be worthwhile when only a single projection is of interest or only a limited number of detection channels is available. In such a scenario, the efficiency could almost be doubled compared to usual 3-mode approach.\\
	An open question is how the presented approach generalizes to higher dimensions. For a d-dimensional system the efficiency of projections with a passive d-mode analyzer scales with $1/d$. In this regime the potential increase in efficiency from using a 2-mode analyzer might be even more substantial.

	\section*{Data availability}
	The data sets generated and/or analysed during the current study are available from the corresponding authors upon reasonable request.
	
	\section*{Acknowledgements}

    We acknowledge funding from the Swiss FNS NCCR programme Quantum Science Technology (QSIT), European Union Horizon 2020 research and innovation program within the Flagship on Quantum Technologies through GA 820445 (QIA) and under the Marie Sk\l{}odowska-Curie program through GA  675662 (QCALL).
    
    We also thank Nicolas Brunner and S\'{e}bastien Designolle from the Universit\'e de Gen\`{e}ve for fruitful discussions and Jean Etesse from the Institut de Physique de Nice for his contributions to the experimental setup. Further, we thank Philippe Goldner and Alban Ferrier from Chimie ParisTech for fruitful discussions and for providing the crystals.
	
	\bibliography{qmcommon}
	
	\newpage
	\section*{Appendix}
	
	\subsection{Finding the parameters for the implementation of PRA}\label{SEC:pra_parameters}
	As explained in section \ref{SEC:theory}, the implementation of the PRA for projecting onto state $\frac{1}{\sqrt{3}}\left(e^{\mathrm{i}\phi_0}\ket{0}+e^{\mathrm{i}\phi_1}\ket{1}+e^{\mathrm{i}\phi_2}\ket{2}\right)$ requires to find the optimal pulse parameters for three partial readouts such that they perform the mapping
	\begin{equation}
		\begin{split}
			\mathbf{U}&\ket{n}_t\ket{s}=\\ &\sqrt{\frac{\eta}{3}}(e^{-\mathrm{i}\phi_2}\ket{n}_t +e^{-\mathrm{i}\phi_1}\ket{n+1}_t+e^{-\mathrm{i}\phi_0}\ket{n+2}_t)\ket{e}\\
			&+\sqrt{1-\eta}(\dots)\ket{s}
		\end{split}
	\end{equation}
	with the highest possible efficiency $\eta$.\\
	A particularly interesting set of projections are the following three bases that are mutually unbiased to each other and the canonical basis (MUBs).
	\begin{equation}\label{MUBs}
		\begin{split}
			&\frac{1}{\sqrt{3}}\left(\ket{0}+\ket{1}+\ket{2}\right) \\
			\mathrm{MUB 1}~~~~
			&\frac{1}{\sqrt{3}}\left(\ket{0}+e^{-\mathrm{i}2/3\pi}\ket{1}+e^{\mathrm{i}2/3\pi}\ket{2}\right) \\
			&\frac{1}{\sqrt{3}}\left(\ket{0}+e^{\mathrm{i}2/3\pi}\ket{1}+e^{-\mathrm{i}2/3\pi}\ket{2}\right) \\
			&~~\\
			&\frac{1}{\sqrt{3}}\left(\ket{0}+\ket{1}+e^{-\mathrm{i}2/3\pi}\ket{2}\right) \\
			\mathrm{MUB 2}~~~~
			&\frac{1}{\sqrt{3}}\left(\ket{0}+e^{-\mathrm{i}2/3\pi}\ket{1}+\ket{2}\right) \\
			&\frac{1}{\sqrt{3}}\left(e^{-\mathrm{i}2/3\pi}\ket{0}+\ket{1}+\ket{2}\right) \\
			&~~\\
			&\frac{1}{\sqrt{3}}\left(\ket{0}+\ket{1}+e^{\mathrm{i}2/3\pi}\ket{2}\right) \\
			\mathrm{MUB 3}~~~~
			&\frac{1}{\sqrt{3}}\left(\ket{0}+e^{\mathrm{i}2/3\pi}\ket{1}+\ket{2}\right) \\
			&\frac{1}{\sqrt{3}}\left(e^{\mathrm{i}2/3\pi}\ket{0}+\ket{1}+\ket{2}\right) \\
		\end{split}
	\end{equation}
	
	We have derived that a time-bin eigenstate entering the readout-based analyzer is mapped onto a superposition with complex amplitudes
	\begin{equation} 
		\begin{split}
			& \zeta_0=-a_0a_1b_2^\star \\
			& \zeta_1=-a_0b_1^\star a_2 + b_0^\star b_1 b_2^\star \\
			& \zeta_2=-b_0^\star a_1a_2
		\end{split}
	\end{equation}
	with $a_i=\sqrt{1-P_i}$ and $b_i=e^{\mathrm{i}\theta_i}\sqrt{P_i}$. For our convenience we factor out a global phase of $\pi$ as such a phase does not affect the state on which we project and yield
	\begin{equation} \label{eq1}
		\begin{split}
			z_0=&a_0a_1b_2^\star\\
			=&\sqrt{(1-P_0)(1-P_1)P_2}e^{\mathrm{-i}\theta_2}\\
			z_1=&a_0b_1^\star a_2 - b_0^\star b_1 b_2^\star\\
			=&\sqrt{(1-P_0)P_1(1-P_2)}e^{\mathrm{-i}\theta_1}\\
			&-\sqrt{P_0P_1P_2}e^{\mathrm{i}(-\theta_0+\theta_1-\theta_2)} \\
			z_2=&b_0^\star a_1a_2\\
			=&\sqrt{P_0(1-P_1)(1-P_2)}e^{\mathrm{-i}\theta_0}
		\end{split}
	\end{equation}
	The first and last bin inherit their phase directly from the first and last partial readout pulse, such that we can immediately conclude that we should set
	\begin{equation}\label{theta13}
		\begin{split}
			& \theta_0=\phi_2 \\
			& \theta_2=\phi_0 \\
		\end{split}
	\end{equation}
	
	Further we demand that the signals in both these bins have the same magnitude.
	\begin{equation} \label{equalouterbins}
		\begin{split}
			& |z_0|^2=|z_2|^2\\
			\implies& \\
			& (1-P_0)(1-P_1)P_2=P_0(1-P_1)(1-P_2)\\
			\implies& \\
			& P_0=P_2
		\end{split}
	\end{equation}
	This leaves $P_0, P_1$ and $\theta_1$ as free parameters. Both the magnitude as well as the phase of the central bin depend on all three of these parameters. We define
	\begin{equation}\label{EQU:ztilde}
		\boxed{
			\begin{split}
				\tilde{z}(P_0,\theta_1)&=z_1/\sqrt{P_1}\\
				&=\sqrt{1-P_0}\sqrt{1-P_2}e^{\mathrm{-i}\theta_1}-\sqrt{P_0}\sqrt{P_2}e^{\mathrm{i}(\theta_1-\theta_0-\theta_2)}\\
				&=(1-P_0)e^{\mathrm{-i}\theta_1}-P_0e^{\mathrm{i}(\theta_1-\phi_0-\phi_2)}
			\end{split}
		}
	\end{equation}
	In order to implement the respective desired projector, we need to ensure that the central bin has the proper phase. 
	\begin{equation}\label{phasecondition}
		\phi_1=-\mathrm{arg}(\tilde{z})
	\end{equation}
	and the same amplitude as the outer bins
	\begin{equation}\label{amplitudecondition}
		|z_1|^2=P_1|\tilde{z}|^2\overset{!}{=}|z_{0}|^2=|z_{2}|^2
	\end{equation}

		Before we move to the individual MUBs, we note that there exist certain classes of analyzers that share the same efficiency $\eta$. For demonstrating this, we first use the global phase freedom to set 
		\begin{equation}\label{EQU:normalform}
		\begin{split}
			\widetilde{\phi}_0&=\phi_0-\phi_1\\
			\widetilde{\phi}_1&=\phi_1-\phi_1=0\\
			\widetilde{\phi}_2&=\phi_2-\phi_1\\
		\end{split}
	\end{equation}
	which is possible without loss of generality. Then, equation \ref{EQU:ztilde} and, consequently, equation \ref{phasecondition} and \ref{amplitudecondition} are only dependent on the $P_i$ and the sum $\widetilde{\phi}_\mathrm{tot}=\widetilde{\phi}_0+\widetilde{\phi}_2$. From this follows that all projectors with the same $\widetilde{\phi}_\mathrm{tot}$ share the same optimal choice for $\theta_1=\widetilde{\theta}_1$, the same optimal transfer probabilities $P_i$ and the same efficiency $\eta$.. Once the initial global phase rotation is undone, this leads to the optimal phase settings
			\begin{equation}\label{EQU:aftertrafo}
		\begin{split}
			&\theta_{0}=\phi_{2}\\
			&\theta_{2}=\phi_{0}\\
			&\theta_1=(\widetilde{\theta}_1+\phi_1)\\
		\end{split}
	\end{equation}
	 We will refer to measurements that are related like this to being of the same efficiency class. It should be noted that all projectors of any given MUB from equation \ref{MUBs} belong to the same efficiency class. That means that in the following we can determine the optimal $\widetilde{\theta}_1$ and $P_i$ for just a single projector from each basis and then determine the optimal parameters for the other two by making use of equation \ref{EQU:aftertrafo}.

	\subsubsection{MUB 1}
	We consider the projection on the first projector of MUB 1. As we set $\theta_{0}=\theta_{2}=0$ in accordance to equation \ref{theta13}, equation \ref{phasecondition} now becomes
	\begin{equation}
		0=\mathrm{arg}(\tilde{z})=\mathrm{arg}\left((1-P_0)e^{\mathrm{i}\theta_1}-P_0e^{-\mathrm{i}\theta_1}\right)
	\end{equation}
	%Depending on whether $P_1<0.5$, $\theta$ is $\pi$ or 0. 
	For $P_0>0.5$ the equation is solved by $\theta_1=\pi$, and $\theta_1=0$ for $P_0<0.5$. In both cases the magnitude is $|\tilde{z}|=|1-2P_0|$. Now we demand that the central and outer bins are of equal magnitude (eq. \ref{amplitudecondition}) and find
	\begin{equation}\label{equal_bin}
		(1-2P_0)^2P_1=(1-P_0)P_1(1-P_0)
	\end{equation}
	This expression is only linear in $P_1$. We can solve for $P_1(P_0)$ and express the magnitude of the time-bins, and therefore $\eta$, purely in terms of $P_0$. 
	\begin{equation}
		\eta=3\cdot(1-2P_0)^2P_1(P_0)
	\end{equation}
	We maximize it so that the resulting analyzer has the highest possible efficiency and conclude
	\begin{equation}
		\begin{split}
			& P_0=(3-\sqrt{3})/6\approx 0.21 \\
			& P_1=1/3 \\
			& \widetilde{\theta}_1=0
		\end{split}
	\end{equation}
	
	\subsubsection{MUB 2 and 3/general case}
	For MUB 2 and 3 we follow a similar route as for MUB 1. Unlike for the latter, no analytical solution has been found. The final optimization step is performed numerically. The approach shown in the following will work for finding PRAs that implement projections onto arbitrary states of the form $\frac{1}{\sqrt{3}}\left(e^{\mathrm{i}\phi_0}\ket{0}+e^{\mathrm{i}\phi_1}\ket{1}+e^{\mathrm{i}\phi_2}\ket{2}\right)$ after transforming them according to equation~\ref{EQU:normalform}.\\
	We once more demand that the phase of the central bin is zero.
	\begin{equation}
	\begin{split}
	\phi_2&=0\overset{!}{=}\mathrm{arg}(\tilde{z}) \\
	\implies&\\
	0&=\mathrm{Im}(\tilde{z})=(1-P_0)\sin(\theta_1)+P_0\sin(\theta_1-\phi_0-\phi_2)\\
	&=(1-P_0)\sin(\theta_1)+P_0\sin(\theta_1-\phi_\mathrm{tot})
	\end{split}
	\end{equation}
	
	Solving for $\theta_1$ gives
	\begin{equation} 
	\theta_1(P_0)=\mathrm{acot}\left(\frac{1/P_0-1+\cos(\phi_{tot})}{\sin(\phi_{tot})}\right)
	\end{equation}
	We insert $\theta_1(P_0)$ into equation \ref{amplitudecondition} and solve for $P_1(P_0)$. Once more we have all ingredients for expressing the total efficiency as a function of $P_0$.
	\begin{equation}
	\eta=3\cdot(1-2P_0)^2P_1(P_0)
	\end{equation}
	We can determine the maximum numerically and find
	\begin{equation} 
	\begin{split}
	& \widetilde{\theta}_1=-0.388 \\
	& P_0=0.2764 \\
	& P_1=0.2857\\
	& \eta=0.4286
	\end{split}
	\end{equation}
	and
	\begin{equation} 
	\begin{split}
	& \widetilde{\theta}_1=0.388 \\
	& P_0=0.2764 \\
	& P_1=0.2857\\
	& \eta=0.4286
	\end{split}
	\end{equation}
	respectively.\\

	\subsection{Optimal basis}
	In order to achieve the highest possible efficiency for the given scheme we have to maximize the magnitude of $\tilde{z}$ from equation \ref{EQU:ztilde}.  This will be the case whenever
	\begin{equation}
		e^{\mathrm{-i}\theta_1}=-e^{\mathrm{i}(\theta_1-\phi_0-\phi_2)}
	\end{equation}
	One possible choice is 
	\begin{equation}\label{conditionbestMUB}
		\begin{split}
			& \phi_{0}=-\phi_{2} \\
			& \theta_{1}=\pi/2
		\end{split}	
	\end{equation}
	The equal bin amplitude condition equation \ref{amplitudecondition} then simply reads:
	\begin{equation}
		P_1=(1-P_0)P_0(1-P_1)
	\end{equation}
	As for the previous bases we solve for $P_1(P_0)$ and subsequently maximize $\eta(P_0)$. We find that the optimal parameter choice and resulting efficiency then reads
	\begin{equation}
		\begin{split}
			& \widetilde{\theta}_1=\pi/2\\
			& P_{0}=0.5\\
			& P_{1}=0.2\\
			& \eta=3/5
		\end{split}	
	\end{equation}

	Finally, we can find two more orthogonal vectors of the same efficiency class and arrive at the basis
	\begin{equation}\label{bestMUB}
		\begin{split}
			& \ket{0}+i\ket{1}+\ket{2} \\
			\mathrm{MUB 1}~~~~
			& e^{-\mathrm{i}2/3\pi}\ket{0}+i\ket{1}+e^{\mathrm{i}2/3\pi}\ket{2} \\
			& e^{\mathrm{i}2/3\pi}\ket{0}+i\ket{1}+e^{-\mathrm{i}2/3\pi}\ket{2} \\
		\end{split}
	\end{equation}

	\subsection{Optimal unitary black box}\label{SEC:blackbox}
	Generally speaking, our device implements the qutrit projections by mapping several time-bins onto one and thereby interfering them. Clearly, such a device cannot be $100\%$ efficient, as it would turn orthogonal inputs into collinear ones, and therefore not be unitary. We can ask what limit on its efficiency we can derive from the unitarity of the device alone. For this purpose, we consider a unitary black box that performs the following mapping.
	\begin{equation} 
		\begin{split}
			\mathbf{U}\ket{n}_t\ket{0}_s = &\sqrt{\frac{\eta}{3}}(\ket{n}_t +e^{\mathrm{i}\phi_1}\ket{n+1}_t+e^{\mathrm{i}\phi_2}\ket{n+2}_t)\ket{0}_s\\
			+&\sqrt{1-\eta}\ket{\psi_n}\\
		\end{split}
	\end{equation}
	$\ket{\psi_n}$ is a unit length vector that contains any contribution that is ending up in the dark port. We can conclude the following.
	\begin{equation}\label{unitary_bound}
		\begin{split}
			0=&|\bra{0_s,2_t}\ket{0_t,0_s}|\\
			=&|\bra{0_s,2_t}\mathbf{U^\dagger U}\ket{0_t,0_s}|\\
			=&(\sqrt{1-\eta}\bra{\psi_2}+\sqrt{\frac{\eta}{3}}\bra{0}_s(\bra{2}_t +e^{-\mathrm{i}\phi_1}\bra{3}_t+e^{-\mathrm{i}\phi_2}\bra{4}_t))\\
			&*(\sqrt{\frac{\eta}{3}}(\ket{0}_t +e^{\mathrm{i}\phi_1}\ket{1}_t+e^{\mathrm{i}\phi_2}\ket{2}_t)\ket{0}_s+\sqrt{1-\eta}\ket{\psi_0})\\
			=&|\frac{\eta}{3}e^{\mathrm{i}\phi_2}+(1-\eta)\bra{\psi_2}\ket{\psi_0}|\\ \iff&\\
			\frac{\eta}{3}&e^{\mathrm{i}(\phi_2-\phi_0)}=-(1-\eta)\bra{\psi_2}\ket{\psi_0}\\
			\implies&\\
			\frac{\eta}{3}&=(1-\eta)|\bra{\psi_2}\ket{\psi_0}|\\
		\end{split}
	\end{equation}
	A bound that we can immediately derive from this is the following
	\begin{equation}
		\begin{split} 
			\frac{\eta}{3}&=(1-\eta)|\bra{\psi_2}\ket{\psi_0}|\\
			&\le(1-\eta)\\
			\implies&\\
			\eta&\le \frac{3}{4}
		\end{split}
	\end{equation}
	But we can derive an even tighter bound by taking the time translation symmetry of our interferometric device into account. Since any delay that the device applies is relative to the time of arrival of the input, the output into the dark ports can be expressed in the following general form.
	\begin{equation}\label{generalortho}
		\ket{\psi_n}=\sum_i a_{i}\ket{n+i}_t\ket{1}_s
	\end{equation}
	
	\subsubsection{Three consecutive non-zero time bins in the dark port}\label{SEC:threebins}
	Before we look at the case of arbitrarily many temporal modes, let us consider the simple case of a device that distributes onto three temporal modes. In this case we can continue from the result of equation \ref{unitary_bound} as follows.
	\begin{equation}\label{threemodesoneport}
		\begin{split} 
			\frac{\eta}{3}=&(1-\eta)|\bra{\psi_2}\ket{\psi_0}|\\
			=&(1-\eta)|\bra{1}_s(a_{2}^*\bra{4}_t+a_{1}^*\bra{3}_t+a_{0}^*\bra{2}_t)\\
			&(a_{0}\ket{0}_t+a_{1}\ket{1}_t+a_{2}\ket{2}_t)\ket{s}|\\
			=&(1-\eta)|a_{0}||a_{2}|\\
			\le&(1-\eta)|a_{0}|\sqrt{1-|a_{0}|^2}\\
			\le&\frac{1-\eta}{2}\\
			\implies&\\
			\eta\le& \frac{3}{5}
		\end{split}
	\end{equation}
	In the fourth line we make use of the fact that $\ket{\psi_n}$ are normalized vectors and in the fifth line we bound $|a_{0}|\sqrt{1-|a_{0}|^2}$ from above by its maximum of $\frac{1}{2}$.
	
	\subsubsection{Arbitrarily many non-zero time bins in the dark port}
	We will see that allowing for arbitrary many modes does not affect this argument. We can show that if we map onto three temporal modes in the bright port, only up to three consecutive temporal modes in the dark port can be populated. 
	
	Assuming that there is a finite amount of temporal modes, there exist a $k$ such that for each input $\ket{n}_t\ket{0}_s$ $a_{n+k}$ is the first non-zero component of the vector $\ket{\psi_n}$, so
	\begin{equation}\label{lowcond}
		\begin{split}
			a_{n+k}&\neq 0\\
			\forall i<n+k: a_{i}&=0
		\end{split}
	\end{equation}
	Further, as there are only finitely many non-zero coefficients, there must be an $l$, such that every coefficient after $a_{n+k+l}$ is guaranteed to be zero.
	\begin{equation}\label{highcond}
		\forall i>n+k+l: a_{i}=0
	\end{equation}
	\\
	For any inputs $\ket{n+l}_t\ket{0}_s$, $\ket{n}_t\ket{0}_s$ with $l>2$ there is no overlap in the bright port. Their overlap in the dark port must therefore also be zero.
	In this scenario, when calculating this overlap we find
	
	\begin{equation}
		\begin{split} 
			0&=\bra{\psi_n}\ket{\psi_{n+l}}\\
			&=\left(\sum_ia_{i}^*\bra{1}_s\bra{n+i}_t\right)\left(\sum_ja_{j}\ket{n+l+j}_t\ket{1}_s\right)\\
			&=\sum_i\sum_ja_{i}^*a_{j}\delta_{n+i,n+l+j}\\
			&=\sum_{i}a_{i+n+l}^*a_{i+n}\\
			&=a_{n+k+l}^*a_{n+k}\\
		\end{split}
	\end{equation}
	$\delta_{n+i,n+l+j}$ in line 3 is a Kronecker delta. From the second to last line we use that for any other term either one or the other coefficient is zero because of condition \ref{lowcond} or condition \ref{highcond}, respectively. \\
	As $a_{n+k}$ is per assumption non-zero, $a_{n+k+l}$ must be zero. Therefore we can now state that only coefficients from $a_{n+k}$ to $a_{n+k+(l-1)}$ can be non-zero. As we reduced the range of possibly non-zero components by one, we can now apply the same proof for the case of $\ket{\psi_{n}}$ and $\ket{\psi_{n+l-1}}$ and will conclude that $a_{n+k+l-1}=0$. We can apply the proof iteratively and reduce the range of possibly non-zero components one by one, right until we reach the case  of $\ket{\psi_{n}}$ and $\ket{\psi_{n+2}}$ where the overlap in the bright port will no longer be zero. Such that we can finally conclude
	\begin{equation}
		\forall l>2: a_{n+k+l}=0
	\end{equation}
	This means there can always be only three consecutive non-zero time bins in the dark port. Even allowing for arbitrarily many time-bins in the dark port, the problem will ultimately be equivalent to the one treated in section~\ref{SEC:threebins}.

	%\bibliography{../../../COMMONBIBFILE/qmcommon}
	
	%\bibliography{../bibliography/qmcommon}

\end{document}